\documentclass[11pt]{article}
\usepackage[ruled, vlined]{algorithm2e}
\usepackage{times}
\usepackage{mathrsfs}
\usepackage{amsfonts}
\usepackage{amssymb}
\usepackage{amsmath}
\usepackage{amsmath,amssymb,color}
\usepackage{graphicx}
\usepackage{epsfig}
\usepackage{setspace}
\usepackage{booktabs}

\makeatletter
\def\ExtendSymbol#1#2#3#4#5{\ext@arrow 0099{\arrowfill@#1#2#3}{#4}{#5}}
\makeatother
\newcommand{\sq}{\hbox{\rlap{$\sqcap$}$\sqcup$}}
\newcommand{\qed}{\hspace*{\fill}\sq}

\begin{document}
\begin{titlepage}

\title{RepTFD: Replay Based Transient Fault Detection}

\author{
   Lei Li, Tianshi Chen, Yunji Chen, Ling Li, and Ruiyang Wu\\
   \small{Institute of Computing Technology, Chinese Academy of Sciences}\\
   \small{Beijing 100190, P. R. China}\\
   \small{(lilei-cpu, chentianshi, cyj, liling, wruiyang)@ict.ac.cn}\\
  }
\date{07/06/2012}
\maketitle  \thispagestyle{empty}

%\IEEEcompsoctitleabstractindextext{
\begin{abstract}
The advances in IC process make future chip multiprocessors (CMPs) more and more vulnerable to transient faults. To detect transient faults, previous core-level schemes provide redundancy for each core separately. As a result, they may leave transient faults in the uncore parts, which consume over 50\% area of a modern CMP, escaped from detection.

This paper proposes RepTFD, the first core-level transient fault detection scheme with 100\% coverage. Instead of providing redundancy for each core separately, RepTFD provides redundancy for a group of cores as a whole. To be specific, it replays the execution of the checked group of cores on a redundant group of cores. Through comparing the execution results between the two groups of cores, all malignant transient faults can be caught. Moreover, RepTFD adopts a novel pending period based record-replay approach, which can greatly reduce the number of execution orders that need to be enforced in the replay-run. Hence, RepTFD brings only 4.76\% performance overhead in comparison to the normal execution without fault-tolerance according to our experiments on the RTL design of an industrial CMP named Godson-3. In addition, RepTFD only consumes about 0.83\% area of Godson-3, while needing only trivial modifications to existing components of Godson-3.

\end{abstract}
\end{titlepage}

%\begin{keywords}
%Transient Fault Detection, Deterministic Replay, Pending Period, Fault Coverage.
%\end{keywords}
%}

%\markboth{Submitted to \textsc{IEEE Transactions on dependable and Secure Computing}}{Li \MakeLowercase{\textit{et al.}}: RepTFD: Replay Based Transient Fault Detection}

%\author{Lei Li, Yunji Chen, Tianshi Chen, Ling Li, and Ruiyang Wu\\\\
%Institute of Computing Technology, Chinese Academy of Sciences\\
%Loongson Technologies Corporation Limited}

%\maketitle

%}

%\IEEEdisplaynotcompsoctitleabstractindextext

%\IEEEpeerreviewmaketitle

%as a result  of higher transistor counts, shrinking device geometries and lowering of operating voltages.
%摘要 150字以内
\section{Introduction} %(2)
\subsection{Motivation}
The operating of a transistor on a chip may meet a transient fault due to various reasons, including environment interference, power supply noise, high-energy particle, and so on.
%According to the estimation of \cite{Shivakumar2002}, the rate of transient fault in a logic gate is expected to reach 1,000 failures in $10^9$ hours (FIT).
Considering the ever increasing number of transistors in commercial chip multiprocessors (CMPs), CMPs are more and more vulnerable to transient faults. Although a transient fault happens only one time and does not appear again in the future execution of a CMP, the error resulted from a transient fault may propagate to other parts of the CMP, causing incorrect instruction results, even system crash.
%Due to the increasing frequency of occurrence and severe negative effect, transient fault detection has become important in the design of commercial CMPs.
%\cite{Shivakumar2002}

A traditional methodology to detect transient faults is to use a redundant core to assist each checked core separately \cite{Gomaa2003,Mukherjee2003,Smolens2006,LaFrieda2007,Shan2011}.
%cyj: 是否漏掉了\cite{Shye} lilei：shye是process level redundancy，直接没有引他
A redundant core has the same program and inputs with the corresponding checked core, thus these two cores should have the same instruction results. Once an instruction on the redundant core has a different result with the corresponding instruction on the checked core, a transient fault is detected. In such a case, both cores should be rolled back to some previous checkpoint using existing checkpointing mechanism \cite{Prvulovic2002,Sorin2002}, and re-execute the program to recover from the fault.

\begin{figure*}[htbp!]
\includegraphics[width=0.45\textwidth]{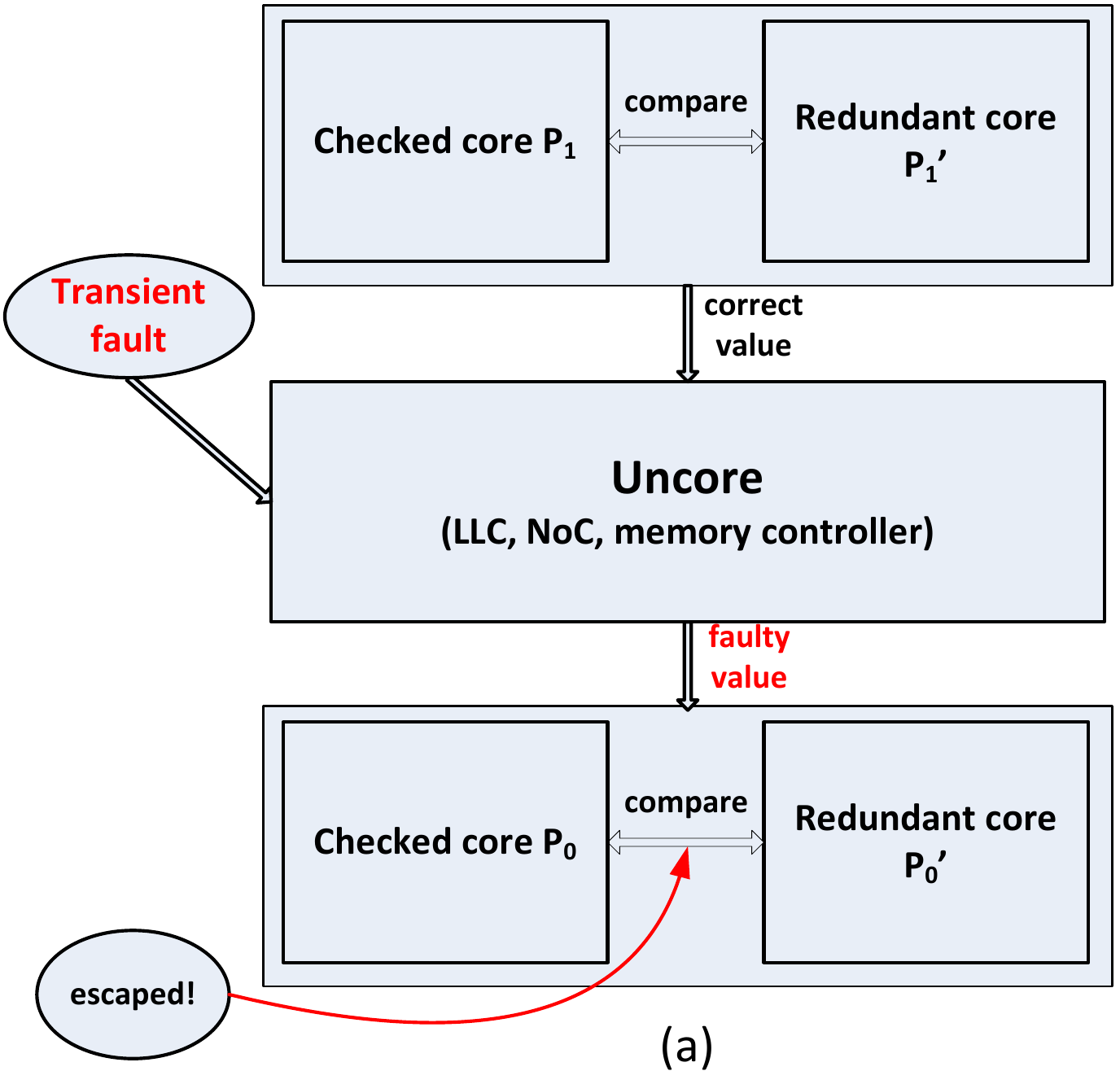}
\includegraphics[width=0.45\textwidth]{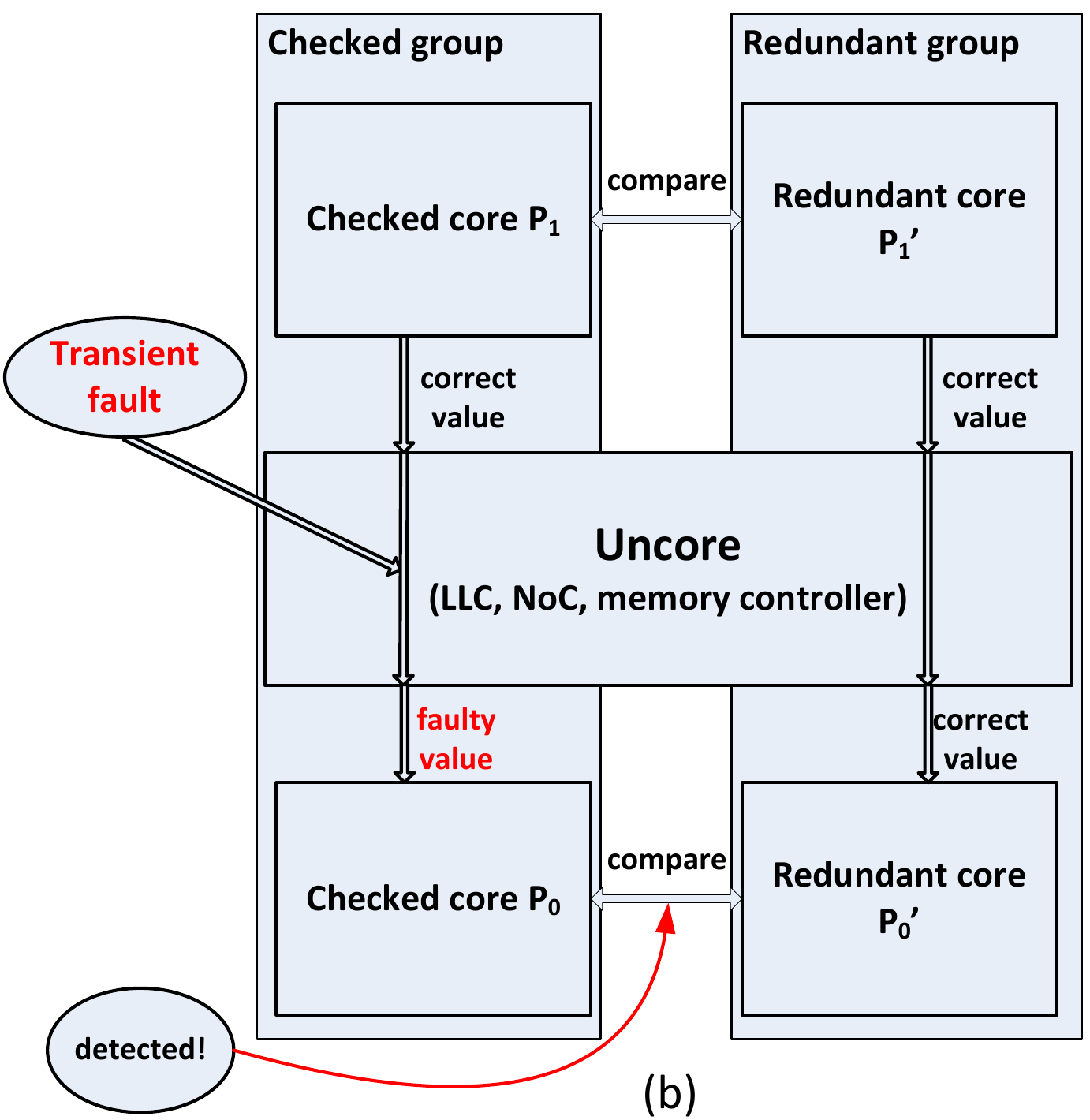}
\caption{Coverage problems of previous schemes. (a) represents previous schemes where a fault in the uncore parts of the chip (last-level cache (LLC), network on chip (NoC), memory controller, and so on) may probably escape from detection. And (b) represents our scheme which can successfully detect all the faults.} \label{fig:3}
\end{figure*}
%cyj: 另外这个图很挫，还不如上一幅。没有突出core group 另外caption说明我们不需要对uncore进行redundancy

Although previous core-level schemes (providing redundancy for each core separately) can effectively detect the transient faults happening in cores, \emph{when coping with parallel programs, they may omit the transient faults happening in the uncore parts of the CMP}, including last-level cache (LLC), network on chip (NoC), memory controller, and so on. An illustrative example can be found in Figure \ref{fig:3}.(a), where core $P_0'$ is the redundancy of core $P_0$, and core $P_1'$ is the redundancy of core $P_1$. Assume that core $P_1$ has correctly stored a value in its local L1 cache. When the correct value stored by core $P_1$ is evicted from core $P_1$ to LLC or memory, it may be corrupted due to a transient fault in the uncore parts, which makes all subsequent loads to the value be faulty. Once core $P_0$ and core $P_0'$ load this faulty value from the uncore, they will produce the \emph{same} incorrect results. Obviously, such a transient fault happening in the uncore parts can be detected through neither the result comparison between $P_0$ and $P_0'$ nor the result comparison between $P_1$ and $P_1'$. Considering that the uncore parts may consume 50-70\% area of a state-of-the-art commercial CMP \cite{Wendel2010,Sawant2011}, there is an urgent need for a scheme which can comprehensively detect transient faults on the whole chip\footnote{Though one can use error correcting codes (ECC) to improve the reliability of the uncore parts, it will remarkably increase the area of the uncore parts (both cache and NoC). Furthermore, ECC cannot cope with many common misbehaviors in the uncore parts (e.g., transfer losing, mistransfers, transfer misordering, incorrect prefetching, and so on).}.

%Furthermore, previous core-level schemes have to pay remarkable efforts to cope with the input incoherence problem between a pair of checked core and redundant core, thus often need remarkable modifications to the critical components of a CMP (e.g., cache coherence, load/store queue, last-level cache, and so on). As a result, they may increase the risk of bug.
%cyj: 想来想去还是删除了。可以问问prvulovic

\subsection{Our Idea}

Previous core-level schemes (which provide redundancy for each core separately) fail to detect all transient faults since the checked core and the corresponding redundant core can load data from a same source. As a result, a transient fault, which affects the common data source of both the checked core and the corresponding redundant core, will escape from detection. \emph{Hence, to get 100\% coverage for transient fault detection, our idea is to provide redundancy for a group of cores as a whole.} As shown in Figure \ref{fig:3}.(b), core $P_0$ and core $P_1$ are treated as the checked group of cores, while core $P_0'$ and core $P_1'$ belong to the redundant group of cores. There is no data dependency between different groups of cores, say, core $P_0$ and core $P_1$ do not rely on any datum produced by the redundant group (core $P_0'$ and core $P_1'$), while core $P_0'$ and core $P_1'$ do not rely on any datum produced by the checked group (core $P_0$ and core $P_1$). When a transient fault occurs, it cannot affect both groups of cores simultaneously. For example, if some transient fault happens in the data transfer from core $P_1$ to core $P_0$, core $P_0'$ will not be affected by this fault. Hence, through comparing the instruction results between core $P_0$ and core $P_0'$, we are able to detect the fault.

\begin{figure}[htbp!]
\includegraphics[width=0.48\textwidth]{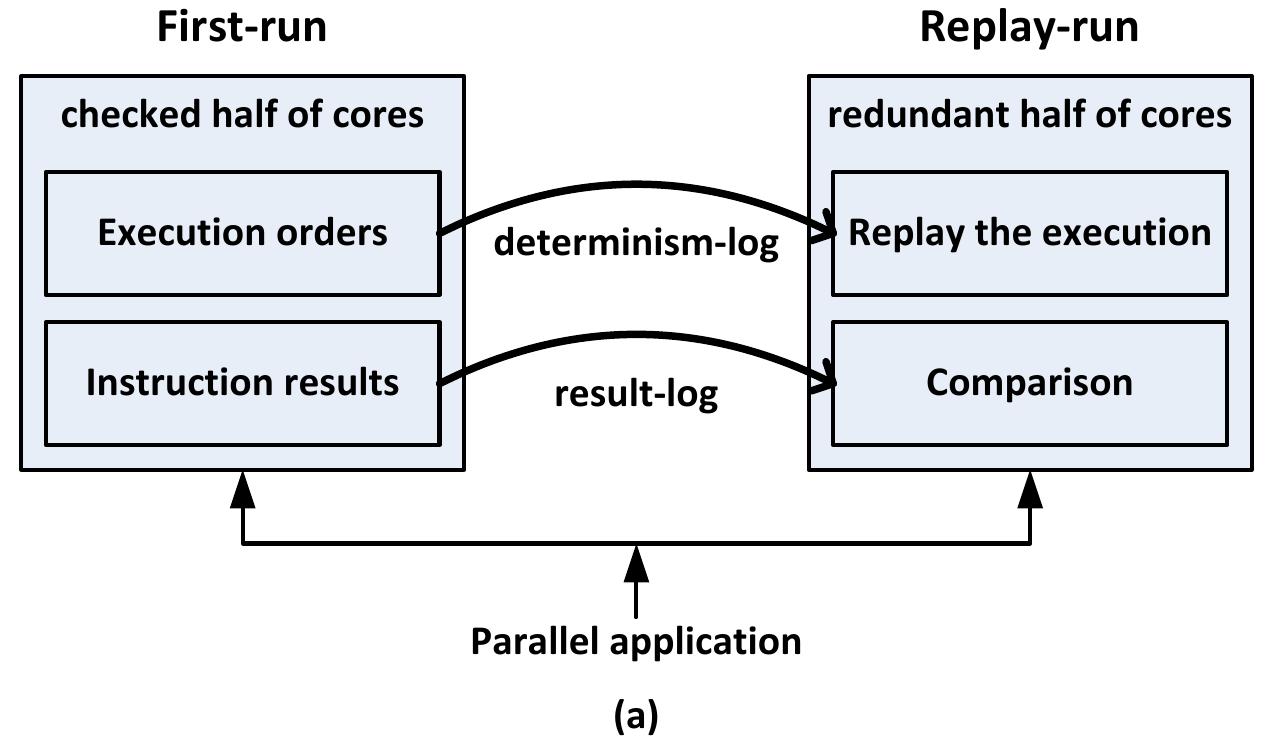}
\includegraphics[width=0.5\textwidth]{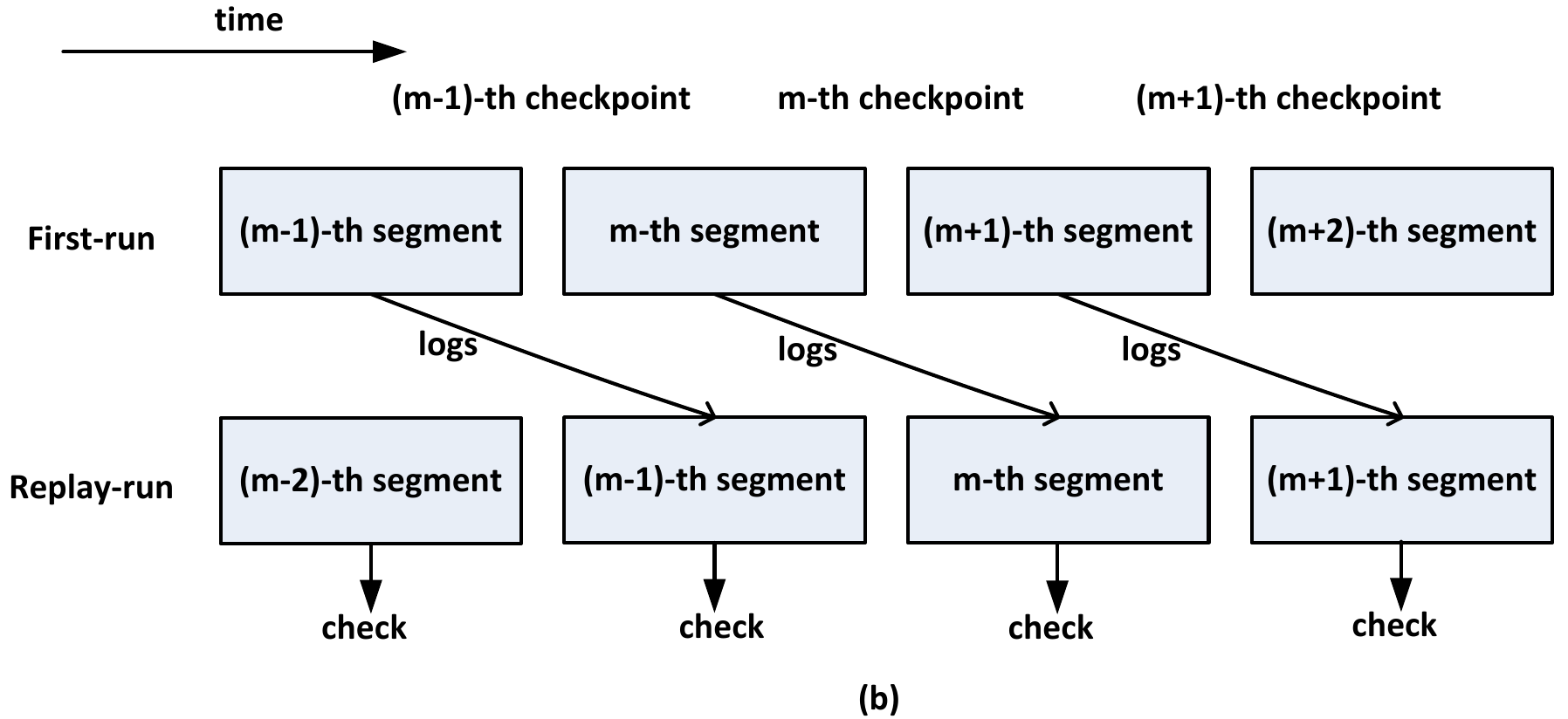}
\caption{RepTFD overview.} \label{fig:5}
\end{figure}
%cyj: caption要加句号。

Based on the above idea, in this paper we propose RepTFD, a transient fault detection scheme based on hardware-assisted deterministic replay\footnote{Deterministic replay aims at guaranteeing two executions of a same program to exhibit the same in the presence of non-deterministic factors. The general workflow of deterministic replay consists of two stages: At the first stage, an execution of the program, called the \emph{first-run}, is carried out, where the non-deterministic factors are continuously recorded as logs. At the second stage, a follow-up execution of the program, called the \emph{replay-run}, recurs the first-run under the guidance of the recorded logs.}. As shown in Figure \ref{fig:5}.(a), RepTFD uses one half of the cores in the CMP as the checked group of cores, and uses the other half of the cores as the redundant group of cores. These two groups of cores execute a same parallel program without data dependency. In the execution on the checked group of cores (first-run), two types of information will be recorded. One is the information of the orders between the memory instructions in the first-run, which is recorded in a \emph{determinism-log} file; the other is the instruction results of the first-run, which is recorded in a \emph{result-log} file. Based on the determinism-log, the first-run can be deterministically replayed on the redundant group of cores. In the meantime, based on the the result-log, the correctness of the first-run can be checked against the replay-run on the redundant group of cores. Since any transient fault can only affect one group of cores, through comparing the results of the corresponding instructions on the two groups, RepTFD can detect malignant transient faults with 100\% coverage (here we say that a transient fault is malignant if it leads to an incorrect instruction result).

Besides the coverage of transient faults, the performance overhead is also a crucial criterion for a transient fault detection scheme. As shown in Figure \ref{fig:5}.(b), the execution of a program with fault-tolerance is divided into many segments using existing checkpoint mechanism. The replay-run of each segment is executed after the execution of the first-run of this segment. Obviously, the overall performance overhead of replay based transient fault detection depends on the slower one of the first-run and the replay-run. Since hardware-assisted deterministic replay approaches can easily achieve negligible slowdown (say, $<1\%$) in the first-run \cite{Chen2010,Voskuilen2010,Hower2008}, the overall performance overhead of RepTFD is determined by the speed of the replay-run. Unfortunately, most existing hardware-assisted deterministic replay approaches focus on reducing the recorded log size to record long execution of parallel program\footnote{Noting that RepTFD only needs to maintain the determinism-log for the last segment of the first-run (about 1 second), the extremely tiny log size achieved by previous deterministic replay approaches has quite limited importance for RepTFD. In practice, the size of RepTFD's determinism-log for a 8-thread application is less than 15 MB when the instruction per cycle (IPC) of each core is 1, which is already satisfactory for transient fault detection.}, thus often cause remarkable slowdown in the replay-run ($>18\%$ to our best knowledge \cite{Chen2010,Voskuilen2010,Montesinos2008,Basu2011}).

Hence, instead of directly incorporating an existing hardware-assisted deterministic replay approaches into RepTFD, we propose a new deterministic replay approach, which has negligible slowdown of the replay-run, based on the following observation: \emph{the more execution orders\footnote{Execution order is a type of logical time order, which exists between two successive conflicting memory instructions. Here we say two memory instruction conflicting, if they access the same memory location and at least one of them is write.} directly recorded, the slower the replay-run is}. For example, if an execution order $u\rightarrow v$ is recorded in the determinism-log, then in the replay-run, before $v$ is executed, we must detect and wait for the completion of $u$, to guarantee the identicalness between the replay-run and the first-run.

Hence, RepTFD achieves efficient replay-run through filtering the execution orders, which can be inferred from \emph{pending period} information, from the determinism-log. Concretely, in the first-run, RepTFD records a relaxed start time and a relaxed end time for each instruction block with a global clock, where the period between these two time points is called the \emph{pending period} of the instruction block. If two instruction blocks have non-overlapping pending periods, they have a \emph{physical time order}. Since execution orders between memory instructions in these two blocks can be inferred from the recorded pending period information (and the resultant physical time order), we need not to directly record them in the determinism-log. In fact, $>99\%$ execution orders are inferrable from physical time orders \cite{Chen2010}. As a consequence, the replay-run of RepTFD is seldom paused to enforce the directly recorded execution orders\footnote{Enforcing the recorded pending periods may also pause the replay-run. However, the pending periods recorded by RepTFD are quite relaxed (i.e., twice of the actual execution time of the instruction block), thus it is rare to pause the replay-run for enforcing the recorded pending periods.}, thus only has 4.76\% slowdown (in comparison to the normal execution without fault-tolerance), according to our experiments over SPLASH2 benchmarks.

%On another hand, a fault in the execution of the $m$-th segment can be detected when replaying this segment, i.e., when the first-run is executing the $(m$+$1)$-th segment. Thus, the detection latency, the time between when a fault occurs and when it is detected, can be controlled about only one checkpoint period.

%relaxed replay constraints of instruction blocks

%强调speed，不是latency

%Moreover, the speed of the replay-run should catch up with that of the first-run in program execution. The first-run may have to wait for the replay-run, which has a performance slowdown with the first-run.

To sum up, this paper makes the following contributions:
\begin{enumerate}

\item{Fault coverage: For parallel workloads, we propose the first core-level transient fault detection scheme with 100\% coverage, i.e., any malignant transient fault that cause an error in the result of an instruction can be detected.}
\item{Performance overhead: Due to the reduction of the execution orders to be enforced in the replay-run, RepTFD has only 4.76\% performance overhead in comparison to the normal execution without fault-tolerance for 16-core CMP (8 checked cores and 8 redundant cores). In addition, RepTFD has the smallest replay slowdown among existing deterministic replay approaches.}
\item{Implementation costs: Previous schemes have to resolve the possible input incoherence between the pair of checked core and redundant core, which incurs non-trivial implementation costs to the CMP. RepTFD elegantly avoids the troublesome input incoherence problem through deterministic replay, and keeps most existing parts of the CMP unmodified. Experiments also show that RepTFD only consumes 0.83\% area of the whole chip. Hence, RepTFD can be easily implemented on a commercial CMP with negligible design, verification, and area costs.}
\end{enumerate}

The rest of this paper is organized as follows: Section \ref{sec:related work} introduces some related investigations. Section \ref{sec:RepTFD} describes the principle and implementation of RepTFD. Section \ref{sec:Experiments} presents some experimental results. Section \ref{sec:conclusion} concludes the whole paper.

\section{Related Work}\label{sec:related work}

\subsection{Core-level Transient Fault Detection}
Researchers have proposed various core-level transient fault detection schemes. These schemes have similar ideas: Using a redundant core to protect each checked core from transient fault separately. When the checked core has different instruction results with that of the corresponding redundant core, a transient fault is detected.

Core-level schemes require a checked core and the corresponding redundant core to have the same inputs \cite{Reis2005,Reis2006,Zhang2010}. For single-threaded applications, this requirement can be straightforwardly satisfied through providing the same program and I/O inputs \cite{Bartlett1986,Slegel1999}. However, for multi-threaded applications, previous schemes meet the so-called input incoherence problem: Even given the same program and I/O inputs, the checked core and the redundant core may still get different results for a same load instruction, if a third-party core stores a new value to the corresponding memory location in the time gap between the load instructions performed by the checked core and the redundant core. To address this input incoherence problem, \cite{Gomaa2003} and \cite{Mukherjee2003} employ a load value queue (LVQ), through which every load value is forwarded from the checked core to the redundant core. However, the LVQ implementation has a high design complexity, especially when the load instructions are executed out-of-orderly.
%Moreover, when the checked core gets a faulty load result, the redundant core will also get the same faulty load result, which may leave transient fault about load instructions undetected.

Instead of forwarding the load results of the checked core to the redundant core, some schemes \cite{Smolens2006,LaFrieda2007,Shan2011} allow both cores to independently access the memory. Reunion \cite{Smolens2006} treats input incoherence similar with transient faults. When an input incoherence is detected, roll-back recovery mechanism should be carried out. As the possibility of an input incoherence is much greater than that of a transient fault, Reunion may bring remarkable performance overhead (up to 250\% for 8 checked cores according to \cite{LaFrieda2007}) for roll-back recovery from input incoherence.
%cyj: 数据来源引用！！！
DCC \cite{LaFrieda2007} maintains a memory access window for each pair of check core and redundant core, to monitor the memory locations which are accessed by both cores. Conflicting store instructions performed by a third-party core to those locations are stalled until both cores have ended their memory instructions. However, the performance overhead is still high (19.2\% on average for 8 checked cores according to \cite{Shan2011}).
%cyj: 数据来源引用！！！
Recently, TDB \cite{Shan2011} is proposed to guarantee the input coherence via cache coherence protocol. Although it has better performance and scalability (with respect to the number of cores) than DCC, its design complexity and risk are quite high, since it needs to redesign a new cache coherence protocol.

Even if the previous core-level transient fault detection schemes bring remarkable modifications to the existing parts of a CMP, they still leave the uncore parts, which consume more than $50\%$ area of a commercial CMP, prone to suffer from transient faults. This is because that these schemes protect each checked core with a redundant core individually. Hence, they allow the checked core and the corresponding redundant core to load value from the same source (say, LLC or memory). If the source itself, contains incorrect values due to some transient fault in the uncore parts (e.g., the transfer from some core's L1 cache to LLC), both the checked core and the redundant core will produce the same incorrect instruction results. As a result, no core-to-core comparison can find such a transient fault in the uncore parts. Furthermore, the previous schemes have to deal with the possible input coherence between the pair of checked core and redundant core, which incurs remarkable design, implementation, and verification costs to the CMP.
%
%\begin{table*}[htbg]
%\caption{Comparisons among RepTFD and some other start-of-art detection schemes.} \centering
%{\footnotesize\begin{tabular}{@{}l|l|l|l@{}}
%\hline
%Schemes & Fault coverage & Main design complexity & Performance overhead \\
%\hline
%Reunion & only core, may bring false positives & modifications on cache controller and processor pipeline& 5\%-250\% for 8 checked cores \\
%\hline
%DCC & only core, no false positive & memory access window, modifications on cache controller & 19.2\% for 8 checked cores \\
%\hline
%RepTFD & the whole chip, no false positive & a $1024\times 27$ CAM in each checked core & 4.76\% for 8 checked cores \\
%\hline
%\end{tabular}\label{table:comparisons}}
%\end{table*}
%cyj: 表项太少，再加几个。另外表第三栏应该是modifications to existing design，突出我们的简单（我们没有在每个核里面加cam!!!!你自己为什么不仔细读自己写的第三章）像我们为了replay改的一堆要加上
%cyj: 表里面是8%，论文里面是4.76%到底多少？？？？认真核对，绝对不能错

Essentially different with the previous core-level schemes, RepTFD treats all checked cores as an entire group, and uses a redundant group of cores, which has the same number of cores with the checked group, to protect the checked group as a whole. Since there is no value dependency between groups (i.e., a checked core and a redundant core will not load a same memory location), any transient fault can affect at most one group of cores, thus can be easily detected. To our best knowledge, RepTFD is the first core-level transient fault detection scheme with 100\% coverage. Furthermore, RepTFD elegantly avoids the troublesome input incoherence problem through deterministic replay, thus does not bring non-trivial modifications to the existing part of a CMP as the previous schemes. Moreover, RepTFD only incurs $0.83\%$ overhead in the chip area and $4.76\%$ performance overhead in comparison to the normal execution without fault-tolerance. To sum up, the effectiveness, elegancy, and efficiency of RepTFD demonstrate its applicability in industry.
%The main design overhead of RepTFD is three auxiliary Bloom filters \cite{Bloom1970} in each checked core to record some execution orders. Table \ref{table:comparisons} shows the comparisons between RepTFD and some other state-of-art schemes. Although the performance overhead of RepTFD is higher than that of TDB, RepTFD has significant superiority on both the fault coverage and the design complexity.

\subsection{Circuit-level Transient Fault Protection}

Besides core-level, a chip can be protected from transient faults at the circuit-level through inserting redundant circuit-level cells (e.g., registers, cells, RAMs). For example, many chips for aerospace engineering adopt Triple Modular Redundancy (TMR), which uses additional two cells to protect one cell from transient faults. Theoretically, any single fault can be detected and corrected with TMR. Moreover, ECC and parity checking, which can protect the memory elements with information redundancy, can cope with transient fault tolerance in cache and memory.

In general, circuit-level protection of transient faults may significantly increase the area of the chip, and slow down the speed. Hence, the usage of circuit-level protection is limited to cache and memory in most commercial CMPs.

\subsection{Protection of the Uncore Parts}
There have also been lots of approaches to make the uncore parts more reliable (especially the NoC). To tolerate incorrect behaviors of the NoC, BulletProof \cite{Constantinides2006} and Vicis \cite{Fick2009} are proposed to detect and recover from the transfer loss. And \cite{Aisopos2011,Puente2004} provide some resilient routing algorithms to re-transfer the networks packets when a fault occurs in the NoC. Recently, \cite{Aisopos2011-2} presents a cache coherence protocol framework which ensures coherence and reliability in the transfers even in the presence of hardware faults of NoC.

However, the above approaches need non-trivial modifications to the existing uncore parts (as well as cores), which may bring additional design and re-verification overheads to the CMP. As a comparison, without any modification on the existing uncore parts, RepTFD can detect transient fault in the uncore parts as well as the cores. Moreover, The fact that a number of approaches have been proposed to protect some uncore parts also demonstrates the urgent need for a fault detection scheme which can cover not only the cores but also the uncore parts.

\subsection{Deterministic Replay}

Deterministic replay aims at guaranteeing two executions of a same program to exhibit the same in the presence of non-deterministic factors. An ideal deterministic replay approach for effective transient fault detection should bring non-remarkable slowdowns to both the first-run and the replay-run. While most of hardware-assisted deterministic replay approaches bring trivial slowdown to the first-run, few of them provides efficient replay-run (partly since they focus on pursuing a small log size, which is unimportant for fault detection) \cite{Chen2010,Voskuilen2010,Lee2010,Lee2011,Veeraraghavan2011,Basu2011}. For example, Karma \cite{Basu2011}, which is a state-of-the-art hardware-assisted deterministic replay approach, still brings about $18-21\%$ slowdown in the replay-run.

\begin{figure*}[htbp!]
\includegraphics[width=1.0\textwidth]{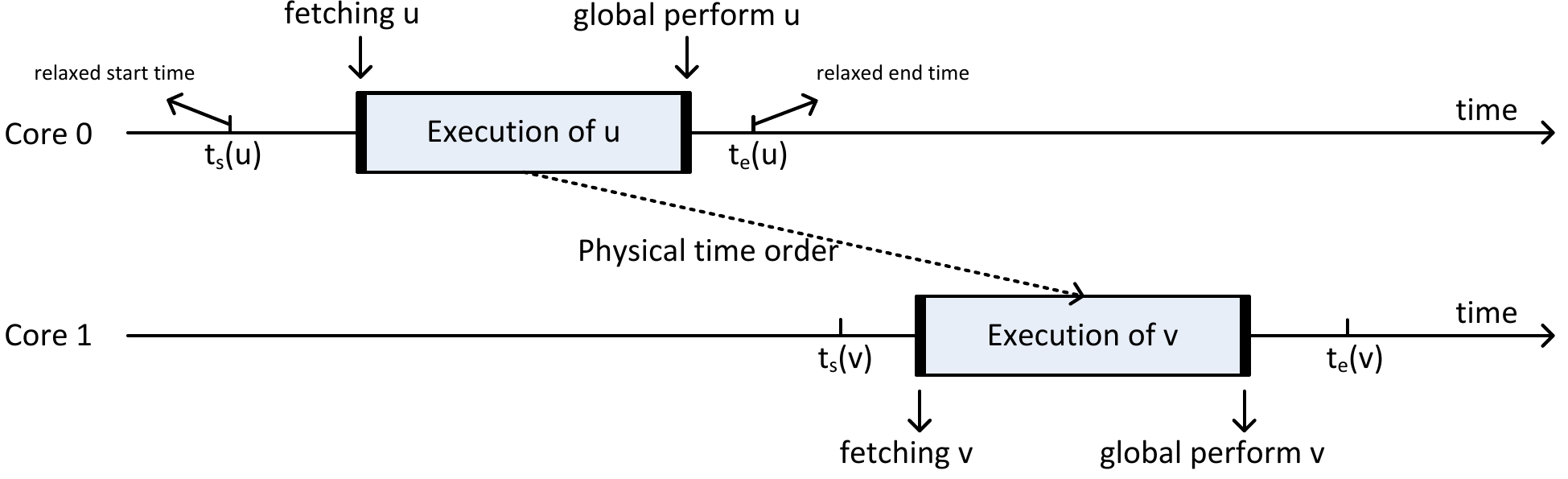}
\caption{An example to illustrate pending period and physical time order. $u$ and $v$ are two memory instructions executed on core0 and core1 respectively. $u$ is before $v$ in physical time order because $u$'s end time of (its pending period period) $t_e(u)$ is earlier than $v$'s start time (of its pending period) $t_s(v)$.} \label{fig:11}
\end{figure*}

To achieve good performance in the replay-run, RepTFD employs pending period based deterministic replay originated from our previous conference paper \cite{Chen2010}. However, to enable efficient replay-run, RepTFD introduced in this journal paper has the following superiorities compared to our conference paper \cite{Chen2010}: 1), RepTFD fixes the execution time of each instruction block (in the record-run), while \cite{Chen2010} fixes the number of instructions in each instruction block. Since the execution time of each block is unbounded in \cite{Chen2010}, some execution orders between blocks with overlapping pending periods cannot be recorded. As a result, \cite{Chen2010} has to enforce these orders through \emph{sequentially} replaying the program, which causes over 700\% slow down in replaying a 8-thread program. In contrast, RepTFD records all non-inferrable execution orders that are between blocks with overlapping pending periods, thus enables replaying the program in \emph{parallel}, which is an essential factor for efficient replay-run. 2), Even if \cite{Chen2010} can replay the program in parallel, its replay-run still has larger overheads to enforce the execution orders non-inferrable from physical time orders, since the unbounded long execution time of block allowed by \cite{Chen2010} brings more non-inferrable execution order (to be enforced in replay-run) than RepTFD.

In summary, RepTFD significantly outperforms our previous conference paper \cite{Chen2010} in the replay speed, which is crucial for fault detection. Moreover, to our best knowledge, the 4.76\% replay slowdown of RepTFD is also the smallest among those of all existing deterministic replay approaches.

%To address this problem, RepTFD adjusts the size of instruction blocks to ensure that the execution time of each instruction block is roughly the same. As a result,  Through monitoring the memory instructions on these blocks with overlapping pending periods, all execution orders can be recorded (detailed in Section \ref{sec:first-run}). As a result, the programs can be replayed on a multi-processor and thus has only 3.85\% speed overhead in the replay-run. To our best knowledge, RepTFD has the smallest performance overhead in the replay-run among existing deterministic replay approaches.

%These committed memory instructions are regarded as an instruction block, and the size of the pending period of each block is two sampling periods, which is 1 time longer than the block's real execution time. As the size of the pending period is always two sampling periods

\section{Implementation of RepTFD}\label{sec:RepTFD}

\subsection{Overview of RepTFD}

RepTFD uses one half of the cores in the CMP as the checked group of cores, and uses the other half of the cores as the redundant group of cores. These two groups of cores execute a same parallel program without data dependency. The first-run (on the checked group of cores) is divided into many segments using existing checkpoint mechanism as shown in Figure \ref{fig:5}.(b). In each segment of the first-run, two log files, which include determinism-log and result-log, are generated. The determinism-log can be used to enforce the replay-run (on the redundant group of cores) behaving the same with the first-run. And the result-log contains the instruction results of the first-run.

After the first-run of the the $m$-th segment, the replay-run of the $m$-th segment starts on the redundant group of cores, and the results in the first-run and the replay-run are compared. Once there is a mismatch between the results of the first-run and the replay-run, a malignant transient fault is detected. Then both the checked group of cores and the redundant group of cores are rolled back to the most recent checkpoint to recover from the transient fault. For example, if a fault is detected when replaying the $m$-th segment, both the first-run and the replay-run should be rolled backed to the $(m$-$1)$-th checkpoint to re-execute the $m$-th segment (noting that the replay-run should wait for the first-run to complete the re-execution of the $m$-th segment first).

%\begin{figure}[htbp!]
%\includegraphics[width=0.5\textwidth]{6}
%\caption{Replay and check the execution for each segment} \label{fig:6}
%\end{figure}

\subsection{First-run}\label{sec:first-run}%（1）
As we have mentioned, in the first-run, two log files, i.e., determinism-log and result-log, should be recorded.

\subsubsection{Determinism-log}

\mbox{}

For deterministic replay, the execution orders among memory instructions should be recorded as determinism-log. To achieve determinism, each recorded execution order (e.g., $u\rightarrow v$) should be enforced in the replay-run with dedicated synchronization: Before $v$ is executed, the replay-run must detect and wait for the completion of $u$. As a result, enforcing the recorded execution orders becomes the main performance overhead of the replay-run \cite{Chen2010,Voskuilen2010,Montesinos2008}. To implement efficient replay-run, RepTFD employs a pending period based recording approach to reduce the number of directly recorded execution orders that need to be enforced in the replay-run.

Before introducing how RepTFD records execution orders, we need to introduce the concepts of pending period and physical time order, which were proposed in our previous investigations \cite{Chen2009,Chen2010}, as preliminaries. As shown in Figure \ref{fig:11}, the pending period of an instruction is a relaxed time interval (denoted as $[t_s,t_e]$) on the global clock in which the instruction is globally performed. For a CMP with a global clock, the pending periods of any two instructions $u$ and $v$ can be compared. If their pending periods do not overlap, there exists a physical time order between these two instructions: If the end time of $u$'s pending period (also denoted as $u$'s end time for brevity) is earlier than the start time of $v$'s pending period (also denoted as $v$'s start time for brevity), then we say that $u$ is before $v$ in physical time order.

In \cite{Chen2009}, Chen \emph{et al.} proved that if an instruction $u$ is before another instruction $v$ in physical time order, $v$ cannot be before $u$ in any kind of logical time order. The reason is quite intuitive: instruction $u$ must have been globally performed before that $v$ begins to execute, thus $v$ is impossible to affect the result of $u$. Therefore, if the physical time order between a pair of two memory instructions is known according to the pending period information, the logical time order (including execution order) between two instructions can be inferred. In fact, most ($>99\%$) execution orders can be inferred from the pending period information according to \cite{Chen2010}.

\begin{figure}[htbp!]
\includegraphics[width=0.9\textwidth]{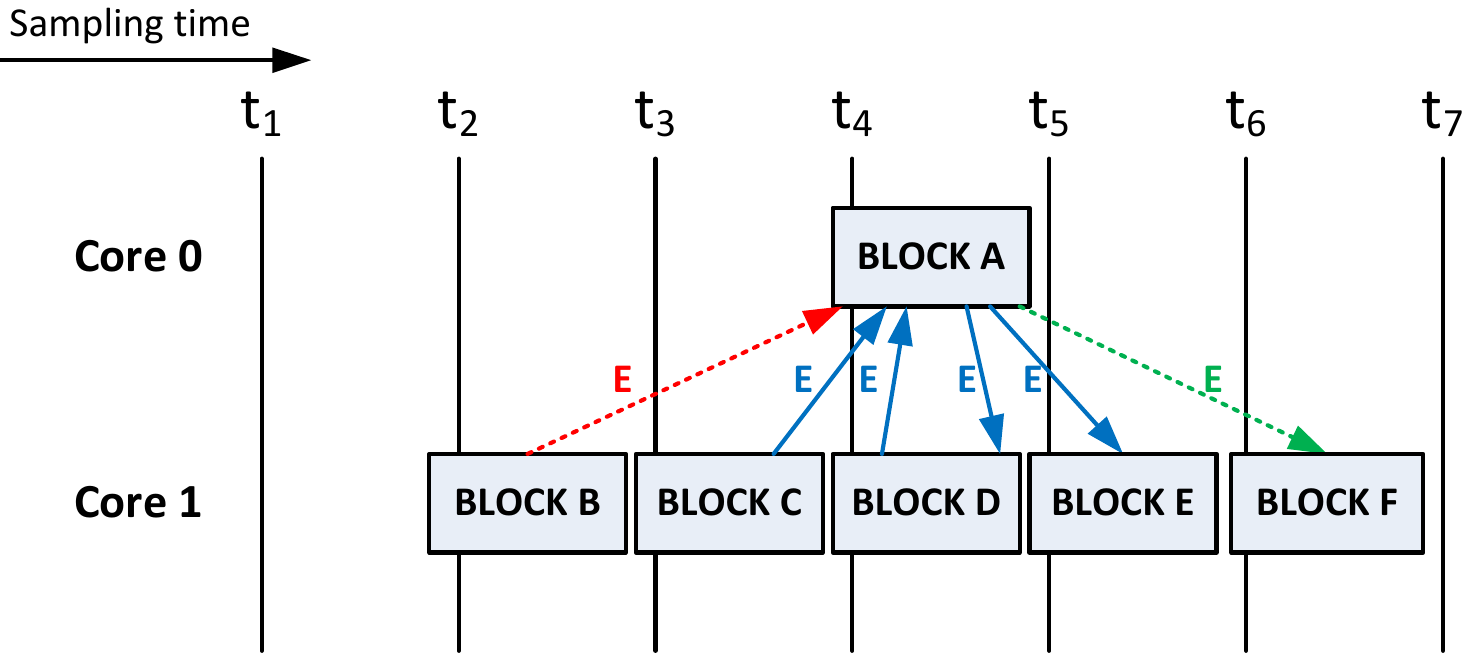}
\caption{Pending period of instruction blocks via sampling. The dashed arrows represent the execution orders inferrable from the pending period information, while the bold arrows represent the execution orders that cannot be inferred from the pending period information.} \label{fig:1}
\end{figure}

\emph{Therefore, instead of directly recording all execution orders, RepTFD records the pending period information and the non-inferrable execution orders (i.e., execution orders that can not be inferred from the pending period information).} To reduce the overhead, RepTFD records the pending period of instruction block instead of individual instruction\footnote{The pending period of an instruction block is the time interval between the start time of the first start instruction in the block, and the end time of the last end instruction in the block.}. Through sampling, each thread is naturally divided into instruction blocks, where the $i$-th instruction block consists of instructions whose global-perform times are between $(i$-$1)$-th and $i$-th samplings. Noting that some instructions in the $i$-th block (whose global-perform time is later than the $(i$-$1)$-th sampling) may begin earlier than the $(i$-$1)$-th sampling. Hence, the pending periods of the $i$-th block is between the $(i$-$2)$-th and $i$-th samplings. \emph{Say, the length of the pending period of each block (two sampling spans) is twice of the actual execution time of the block (one sampling span).} As shown in Figure \ref{fig:1}, instructions in block $A$ are those instructions on core0, whose global-perform times are in the region of $[t_4, t_5]$, while the pending period of block $A$ is $[t_3, t_5]$.

Figure \ref{fig:1} provides an illustrative example about which execution orders should be recorded. Block $A$ is an instruction block executed on core0, and block $B$, $C$, $D$, $E$, $F$ are five consecutive instruction blocks executed on core1. The pending periods of block $A, B, C, D, E, F$ are $[t_3,t_5]$, $[t_1,t_3]$, $[t_2,t_4]$, $[t_3, t_5]$, $[t_4, t_6]$, $[t_5, t_7]$ respectively. With pending period information, the red dashed execution orders, i.e., the execution orders between memory instructions in block $A$ and memory instructions in block $B$ need not to be recorded. These execution orders are inferrable, because memory instructions in block $B$ have globally performed at sampling time $t_3$ while memory instructions in block $A$ have not started. Similarly, all the green dashed execution orders are also inferrable. Actually, only the execution orders between memory instructions in block $A$ and memory instructions in block $C$, $D$ and $E$ (the bold execution orders) are non-inferrable and thus should be recorded.

Concretely, to record the pending period information in the determinism-log, RepTFD employs a PC-sampling technique. At a sampling time $T$, instructions that have committed out of the instruction window should have a end time before $T$. Meanwhile, instructions that have not entered the instruction window should have a start time after $T$. Thus, we can obtain the pending period of each instruction by observing the instructions in the instruction window at every sampling time.

%After each sampling period, the value in the counter is exported to be recorded in the logs and then re-initialized. When reading the values in the logs, the pending period of each instruction block can be obtained. Notably, the size of each instruction block is no more than the size of each sampling period\footnote{In most CMPs, each core can execute at most one memory instruction per clock cycle.}.

%To record the pending period information in the determinism-log, we only need to record the the detail number of memory instructions in each block. With the pending period of each instruction block, we can know whether there is a physical time order between any pair of blocks through comparing their pending periods. As mentioned, if two instruction blocks have known physical time order, any execution order between instructions in the two blocks is inferrable, thus RepTFD needs not to directly record these execution orders.

To record these non-inferrable execution orders (i.e., execution orders that can not be inferred from the pending period information), RepTFD employs a CAM for each core to save the accessed addresses of memory instructions in the last block and the current block of the core. If a local L1 miss occurs when executing instruction $u$, the address of $u$ is searched in the CAMs for other cores to check whether there is a conflicting memory instruction. If $u$ hits in the accessed address of some instruction $v$ saved in the CAMs, the information of $u$ and the information of $v$ should be both recorded as an non-inferrable execution order $v\rightarrow u$. Take Figure \ref{fig:1} for instance. At the time between $t_4$ and $t_5$, core0 is executing block $A$ and core1 is executing block $D$. Memory operations in block $C$ and $D$ should be stored in the CAM for core1. If a L1 miss occurs on core0 when executing an instruction in block $A$, it searches the CAM for core1 to see whether there is a conflicting memory instruction. In this manner, all non-inferrable execution orders from instructions in block $C$ or $D$ to instructions in block $A$ can be recorded. Similarly, all non-inferrable execution orders from instructions in block $A$ to instructions in block $D$ or $E$ can also be recorded when core1 checks the CAM of core0. As a result, all non-inferrable execution orders, which lie between blocks with overlapping pending periods, are recorded.

%As there are at most two instruction blocks stored in the CAM, the size of the CAM should be twice of the sampling period.

%\footnote{All write-after-write and read-after-write style execution orders are recorded, while write-after-read style execution orders are not recorded since they can be inferred from the above two style execution orders.}

Notably, the length of the sampling span is closely related to the replay speed. As shown in Figure \ref{fig:1}, the pending period of an instruction block (two sampling spans) is the relaxation of the accurate execution time of the block (one sampling span). Hence, the higher sampling frequency, the more accurate the pending period is, and the less relaxation for each block. As a result, for a shorter sampling span, the replay-run needs to pay more efforts to enforce physical time orders. On the other hand, a longer sampling span will cause lesser inferrable execution orders, which means the replay-run needs to pay more efforts to enforce non-inferrable execution orders. In practice, RepTFD configures the sampling span to be 512 clock cycles as a tradeoff between the efforts to enforce physical time orders and to enforce non-inferrable execution orders in the replay-run. %XX和XX作用他们的大小，以尽量减少replay-run的等待时间。

\subsubsection{Result-log}

\mbox{}

The result-log needs to record the instruction results in the first-run. Without lost of generality, assume that each instruction needs to modify a 32-bit register or a 32-bit memory location. However, recording 32 bits for each instruction may need hundreds of GByte per second, which is unpractical for state-of-the-art I/O and memory interfaces. To reduce the size of the result-log, we utilize a technique called CheckSum, which can significantly reduce the amount of data with lossy compression.

\begin{algorithm}[!ht]
\caption{CheckSum Technique}
\label{alg:CheckSum}
{
module CheckSum;\; \Begin{
input [31:0] $result$;$\quad$/*result of an instruction*/\;
input $new\_commit$;$\quad$/*commit an instruction*/\;
input [63:0] $commit\_instructions$;\;
output reg [31:0] $checksum$;\;
output $out\_valid$; $\quad$/*valid when outputs the checksum*/\;
\quad\;
always@ (posedge clock)\;
\If{(reset $||$ out\_valid)}
    {$out\_valid$ $<$= 1'b0;\;
    $checksum$ $<$= 31'b0;\;\quad /*initialization*/\;}
\Else
{
\If{(new\_commit)}{
    $checksum$ $<$= $checksum$ XOR $result$;\;}
\If{($commit\_instructions$\%1024==0)}{
    $out\_valid$ $<$= 1'b1;\;
    $export$($checksum$, $result\_log$);\;
}
}
}
}

\end{algorithm}

As shown in Algorithm \ref{alg:CheckSum}, RepTFD employs a 32-bit checksum register and a few gates for each core to summarize the instruction results on the core. The value of each checksum register is initialized to 0 at reset. When committing an instruction, the new value of the checksum register is changed to the the result of the instruction XOR the old value of the checksum register. For every 1024 instructions, the $out\_valid$ register is set to be $1'b1$. At such time, the current value of the checksum register is exported to the result-log, and then re-initialized. Obviously, the size of the CheckSum result-log is quite small: it only consumes 4 byte per kilo instructions. Furthermore, besides CheckSum, other lossy compression approaches can also be applied to reduce the size of the result-log.

\subsection{Replay-Run}\label{sec:replayrun}%(1)

In the replay-run, the same program is replayed on another half cores of the CMP (the redundant group of cores) according to the determinism-log. In addition, the instruction results of the replay-run should be compared dynamically with the result-log recorded in the first-run.

\subsubsection{How to Replay}

\mbox{}

For faithfully replaying, two types of information recorded in the determinism-log should be enforced in the replay-run. One is the pending period information (as well as the resultant physical time orders). The other is the non-inferrable execution orders.

To enforce the physical time orders, in the replay-run, an instruction block $A$ cannot be executed unless all instruction blocks before $A$ in physical time order have ended their executions. Instead of analyzing the pending period information to get detailed physical time orders, we utilize a ``grant'' array in the replay-run to enable efficient replay. Each element of the grant array represents the state of a sampling time. Concretely, the array element of each sampling time $s$ has a value initialized to $0$. This value is increased by $1$ when a thread finishes all of its instruction blocks whose pending periods end no later than $s$. Any instruction block, whose pending period starts no earlier than $s$ in the first-run, cannot be executed in the replay-run, unless the value of array element corresponding to $s$ has been increased to $p$ (the number of threads), which indicates that all threads have granted the execution of instruction block starting later than $s$. In this way, all physical time orders are guaranteed in the replay-run with negligible costs.

%\begin{figure*}[htbp!]
%\includegraphics[width=1\textwidth]{8}
%\caption{Actions on the \textbf{three registers} when a core ends a block and starts a new one.} \label{fig:8}
%\end{figure*}

In practice, three registers should be assigned for each core to implement the above idea. The first register, denoted by $next\_start$, is the start time of the next block's pending period. The second register, denoted by $curr\_end$, is the end time of the current block's pending period. And the third register, denoted by $next\_end$, is the end time of the next block's pending period. %Specifically, $next\_end = curr\_end+1=next\_start+2$ since the pending period of each block is twice of the sampling period.

Algorithm $\ref{alg:replay}$ is the detailed replay algorithm to enforce the physical time orders. $grant(i)$ is the aforementioned grant array, which represents how many cores have finished their instruction blocks whose pending periods end no later than $s$. When an instruction block ends, the grant values of sampling time between $curr\_end$ and $next\_end$ are increased by $1$, since the corresponding core have finished all of its instruction blocks whose pending periods end no later than these periods. On the other hand, when a new instruction block is about to start, the grant value of its start time, $grant(next\_start)$ in the algorithm, should be checked. This block can be executed only if the grant value equals to the total number of cores $p$. Otherwise, it should wait and stall its execution.
%cyj: 上一段还是乱，再整整。特别是上一段中一句Because...简直就是没头没脑。because的结果也没有

%The concrete replay of pending period and physical time order is illustrated in Figure \ref{fig:8}.  \textbf{ If the $m$-th block is current executing instruction block on core $0$,  The execution of the $(m$+$1)$-th block should wait for the tick of period $register1$. And when the $m$-th block ends its execution, core $0$ should inform all the ticks with periods before $register2$. In the mean time, the value of the three registers should also be updated.}

\begin{algorithm}[h!]
\caption{Replay for Physical Time Orders}
\label{alg:replay}
{
module replay\_p;\; \Begin{
reg [7:0] next\_start, [7:0] curr\_end, [7:0] next\_end;\;$\quad$ /*the three mentioned registers*/\;
reg [7:0] next\_start\_new, [7:0] next\_end\_new;\;$\quad$ /*to update the value when a block ends or starts*/\;

\If{(end\_a\_block)}{
    \For{($i =$ curr\_end; $i<$next\_end; $i++$)}{
     $grant$(i)$++$;\;
   }
   curr\_end = next\_end;\;
   next\_end = next\_end\_new;\;

}
\If {(start\_a\_block)}{
   \If {($grant(next\_start) == p$)}
       {$\quad$$\quad$/*p is the number of cores*/\;\;
       start\_a\_new\_block;\;
         next\_start = next\_start\_new;\;
}

    \Else
        {pause\_the\_progress;}\;
}

}
}

\end{algorithm}

%To be aware of when the core starts and ends the execution of an instruction block, for each core there should be a committed inst-counter and a fetched inst-counter to record the number of committed memory instructions and fetched memory instructions respectively. Through monitoring the values in this two inst counters, the precise moment to update the value in the three registers can be obtained.
%
%Noting that in most commercial CMPs, the fetching and committing of memory instructions are both sequentially. As a consequence, we only need to check the grant value when fetching the first instruction in an instruction block, and send an allow request (to other cores) when committing the last instruction in an instruction block.
%cyj:上一段依然不能让人满意。allow request是请求。你是在请求别人allow吗？似乎又不是，似乎你是在allow别人。直接叫allow signal for xxx?

%Considering that the array $grant$ assigns a 4-bit register for each sampling time, the size of the array can be limited through dynamically freeing unnecessary grant values.

Besides the execution orders indirectly guaranteed by the physical time orders, the directly recorded non-inferrable execution orders should also be enforced in the replay-run. Consider an execution order $u\rightarrow v$ whose correlative instructions $u$ and $v$ are executed on core0 and core1 respectively. In the replay-run, core1 cannot execute instruction $v$ until $u$ has been committed by core0. Hence, for each core, RepTFD inserts a dedicated execution order buffer to import execution order information\footnote{According to our experiments, for every 10,000 instructions, there is only less than 1 non-inferrable execution order directly recorded in the determinism-log on average. Thus, only a tiny buffer (e.g., 32 byte) is enough to buff the recorded execution orders.}, and a single bit to pause the progress of the core. With the information in the execution order buffer, core1, which executes instruction $v$, knows that it should wait for the completion of $u$ first. If $u$ has not been completed, the pause bit of core1 is set, thus core1 pauses its own progress. When $u$ has been accomplished by core0, core0 will acknowledge core1. As a result, core1's pause bit is cleared, and $v$ can be executed by core1. In this way, all execution orders directly recorded in the determinism-log can be guaranteed.

\begin{figure}[htbp!]
\includegraphics[width=1.0\textwidth]{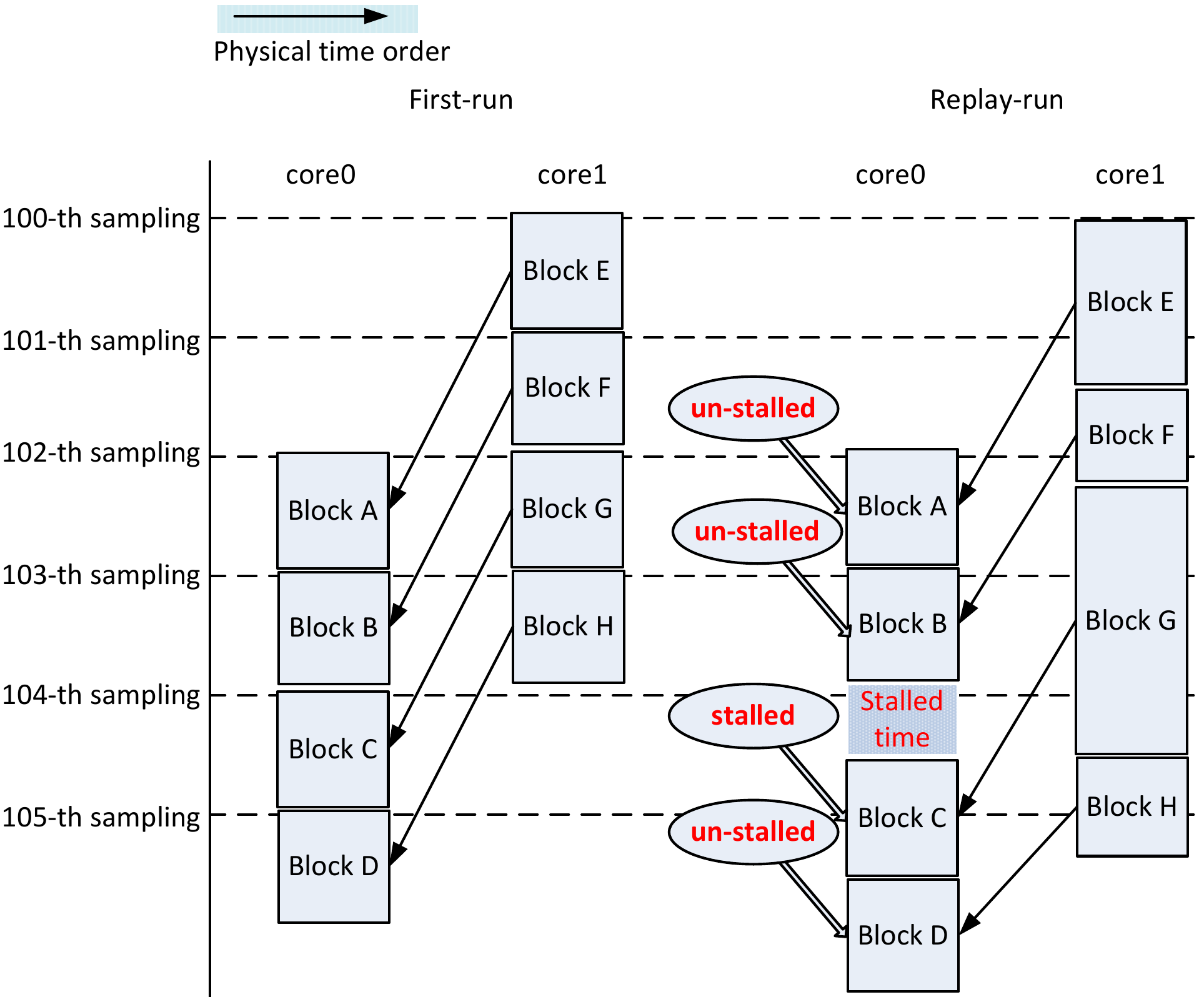}
\caption{An example to illustrate the performance loss to enforce the physical time orders. Each block takes one sampling span (512 cycles) to execute in the record-run, while their pending periods are two sampling spans (1024 cycles). In the replay-run, only block C will be stalled to enforce the physical time order $G\rightarrow C$ since the end time of block G is delayed 512+ clock cycles.} \label{fig:18}
\end{figure}

As mentioned, the major concern about the replay-run is the speed overhead, which is produced when an instruction is stalled to enforce some order (either physical time order or non-inferrable execution order) with other instruction. To enforce the physical time orders in the replay-run, RepTFD only needs to guarantee that each block can be executed within the given pending period. Recall that the length of the pending period for each instruction block is two sampling spans, while the actual execution time of each block is only one sampling span. As a result, only a few blocks should be stalled in the replay-run to enforce the recorded pending periods and physical time orders. Meanwhile, the overhead for enforcing the directly recorded non-inferrable execution orders is also low, since the amount of non-inferrable execution orders is quite few (only $<1\%$ execution orders are non-inferrable from the physical time orders). To sum up, RepTFD will not bring remarkable slowdown in the replay-run.
%cyj: 还有xx\%!!!你的数据到底是怎么搞的？？？

\begin{figure*}[htbp!]
\includegraphics[width=0.85\textwidth]{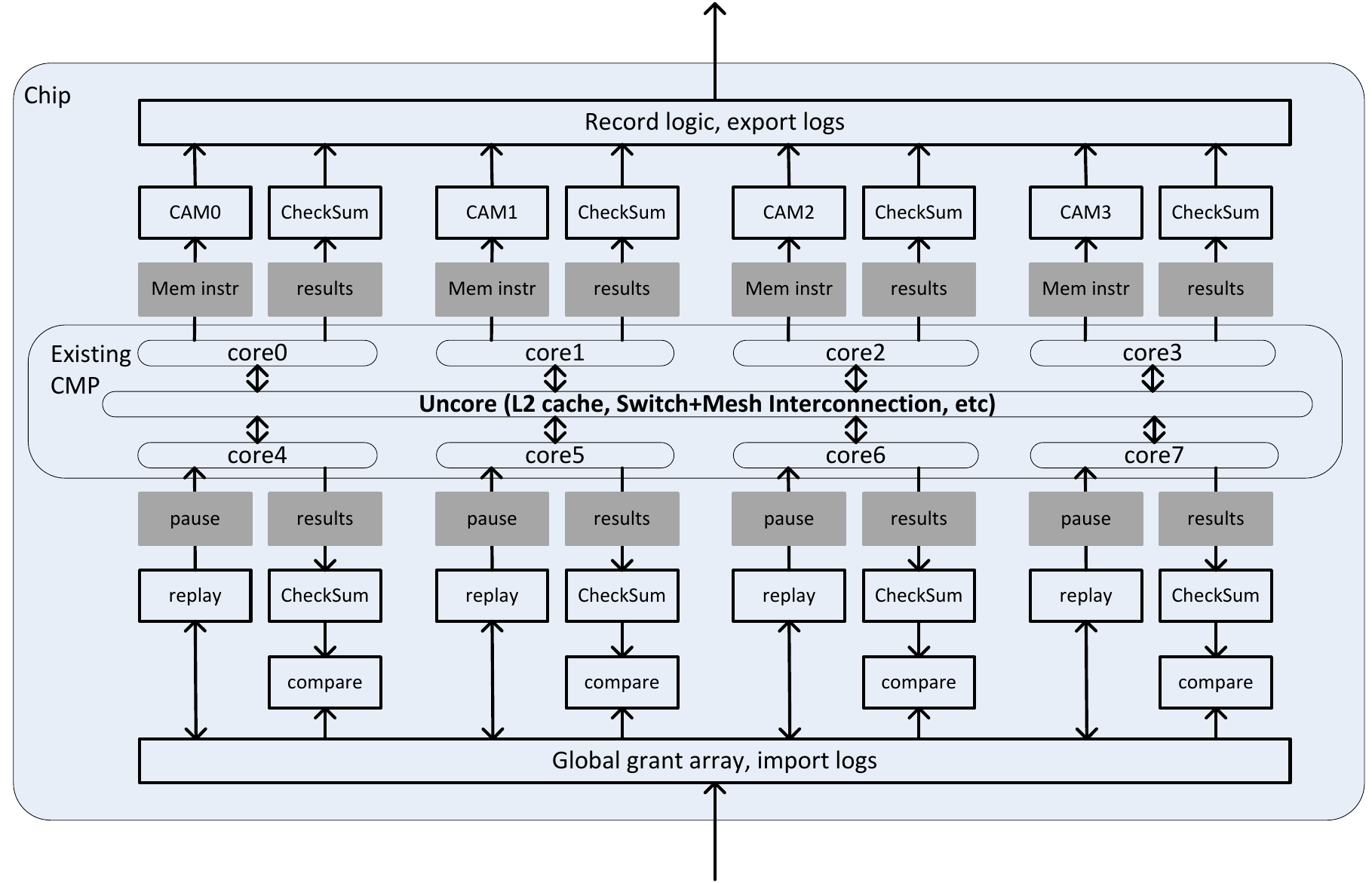}
\caption{Hardware Implementation for RepTFD. core0 -- core3 are the checked group of cores. And core4 -- core7 are the redundant group of cores. } \label{fig:13}
\end{figure*}

Take Figure \ref{fig:18} for example to illustrate RepTFD's low performance overhead of replay-run. Block A, B, C, D are four consecutive blocks executed on core0, and Block E, F, G, H are four consecutive blocks executed on core1. The actual execution time of each block in the first-run is one sampling span (512 cycles). The arrows represent the physical time orders between these blocks in the first-run. For example, block E is before block A in physical time order, since the pending periods of block A and block E are from the 101-st sampling to the 103-rd sampling, and from the 99-th sampling to the 101-st sampling respectively. Suppose each block on core0 has the same execution time in the replay-run as in the first-run. In the replay-run, for block A, B, and D, their executions are not stalled, since block E, F, and H are accomplished before block A, B, and D start respectively. The only exception is block C, which is stalled since the end time of block G is delayed for more than 512 clock cycles (longer than one sampling span). According to our experiments, we found that only $14.17\%$ instruction blocks on average are stalled to enforce the physical time orders, and the average stalled time is only $21.75$ clock cycles, which brings very low overhead to the performance of the replay-run.
%cyj: 上面这段图重画加入execution order重写
%lilei: 这个地方很难把execution order加入图里面，execution order可以由physical order infer感觉在上一个figure说清楚了

\subsubsection{How to Check Results}

\mbox{}

The results in the replay-run are expected to be the same as those in the first-run. Therefore, through comparing the results of two runs, transient faults can be detected. RepTFD adopts dynamical comparison, which compares the execution results via dynamically importing the values in the result-log.

Specifically, the results in the replay-run should be lossily compressed with the same algorithm as in the first-run (c.f. in Algorithm \ref{alg:CheckSum}). Hence, each redundant core for the replay-run should also generate a checksum result for every 1024 instructions. Once the checksum result is generated in the replay-run, it should be compared with the corresponding checksum result generated in the first-run (which is saved in the result-log). Obviously, such comparison has only trivial hardware costs, and does not affect the performance of the replay-run.

It is worth noting that, due to lossy compression there may be a rare case that the first-run and the replay-run have the same checksum result of the instruction results, but their instruction results are different. In such a rare case, a transient fault may not be detected immediately. Such theoretical problem is met by many previous transient fault detection schemes \cite{Smolens2006,LaFrieda2007,Shan2011}. However, a transient fault, even not detected immediately, will be finally detected by comparing the results of future instructions affected by the fault \cite{Nomura2011}.

\subsection{Hardware Supports of RepTFD}%(1.5)

%不要再讲LReplay了，直接将RepTFD有哪些模块，每个模块干什么，如何实现。画出图来看图说话。

%The implementation of RepTFD leverages the Lreplay work in \cite{Chen2010} to detect transient faults. Just like LReplay,  In this section, the implementation of LReplay is brief introduced, then the hardware extensions from LReplay to RepTFD is proposed in detail.

In this subsection, we present hardware modifications on the RTL design of an industrial CMP, which is named Godson-3, to implement RepTFD. For the existing design of Godson-3 (whose architecture features can be found in Table \ref{tab:Godson-3}), RepTFD only trivially modifies each core as follow 1) adding a 64-bit registers for each core to count the number of committed memory instructions, 2) implementing a pause bit for each core to pause its progress to guarantee orders in the replay-run. Most existing components of Godson-3 (including L2 cache, memory controller, cache coherence protocol, and switch+mesh interconnection) remain unmodified.

\begin{table*}[htbg] \centering \caption{Architecture Features of Godson-3.} \label{tab:Godson-3}
{\footnotesize\begin{tabular}{@{}r|l@{}} \hline
Number of Processor Cores & $2-16$ cores\\
\hline
Frequency   & $1.0-1.5$ GHz\\
\hline
Network     & Crossbar+Mesh\\
\hline
Pipeline    & Four-issue, nine-stage, out-of-order\\
\hline
Functional Unit & Two 64-bit fix-point units, two 64-bit floating-point units, one 128-bit memory unit\\
\hline
Register File & 32 logical registers and 64 physical registers for fix-point and floating-point respectively\\
\hline
L1 dcache   & Private, 4 way, 32KB, writeback, 32B per line, load-to-use 3(fix)/4(ft) cycles\\
\hline
L1 icache   & Private, 4 way, 64KB, 32B per line\\
\hline
L2 Cache    & Unified address, Share, 4-16 bank, 512K-1MB per bank, 4 way, 32B per line, 30-35 cycles latency\\
\hline
Memory      & 1-4 DDR2/3 controller, $>1$GB per controller, $333$MHz frequency, $\sim 120$ cycles latency\\
\hline
Cache Coherence & Directory-based MSI protocol\\
\hline
\end{tabular}}
\end{table*}

Figure \ref{fig:13} shows the detailed implementation of RepTFD. core0 -- core3 are the checked group of cores. And core4 -- core7 are the redundant group of cores. For the top half of Figure \ref{fig:13}, the checked group of cores needs the following hardware supports to generate the determinism-log and the result-log. (Note that most of these hardware supports are decoupled from the existing components of Godson-3, thus do not bring modifications to these existing components.)

\begin{enumerate}
\item{A counter for each checked core, which copes with PC-sampling to count the committed memory instructions at every sampling time to record the pending period information. As PC-sampling is already supported by most commercial CMPs, recording the pending period information just needs very few cost.
}
\item{A CAM for each checked core, which stores the memory operations in the last executed block and the current executing block. The size of the CAM is $1024\times 27$, because there are at most $1024$ instructions (two sampling spans) to be stored and each memory instruction needs $27$ bits (including the type, address, counter, cache hit).
%cyj: cash hit是什么, lilei：cache hit, 已修正
}
\item{Some logic to record the non-inferrable execution orders: When a L1 miss occurs, the address of the corresponding instruction is searched in the CAMs for other cores. If it hits, the core number and inst counter of this instruction and the hit instruction in the CAM should be both recorded.}

\item{A CheckSum module for each checked core to record the result-log. This module receives the instruction results from the core and then processes the results using lossy compression. As shown in Algorithm \ref{alg:CheckSum}, this module consumes less than 100 bit registers.}

\item{Some logic to export the logs out of the chip.}
\end{enumerate}

As shown in the bottom half of Figure \ref{fig:13}, the following hardware supports are needed by the the redundant group of cores to replay the determinism-log and check the result-log.

\begin{enumerate}
\item{A pause bit for each core to pause its progress to guarantee orders in the replay-run.}

\item{A replay module for each redundant core to enforce both the physical time orders and the non-inferrable execution orders. When an instruction should wait to enforce some order, the replay module sends a signal to pause the progress of the corresponding core. To enforce the physical time orders, several hundreds bit registers are consumed for each core to implement Algorithm \ref{alg:replay}. In the meantime, only a 16-entry buffer is required to enforce the non-inferrable execution orders.}

\item{A global grant array for the redundant group of cores to enforce the physical time orders. This array cooperates with the replay module for each redundant core according to Algorithm \ref{alg:replay}.}
%cyj: 上一句没有信息量。改成xx is needed by xx to do xx.

\item{A CheckSum module for each redundant core to check the instruction results of the first-run.}

\item{Some logic to import the recorded logs into the chip.}

\end{enumerate}

To sum up, the main design overhead of RepTFD is a $1024\times 27$ CAM per checked core (it is also feasible to replace the CAM with space-efficient Bloom filter), thus brings only a few costs to the chip area. Moreover, RepTFD has a very low design complexity because most components of the existing CMP design remain unmodified.

\section{Experimental Results}\label{sec:Experiments}%(2)\label{sec:related work}

%The replay-run incurs additional performance overhead due to enforcing both the physical time orders and the non-inferrable execution orders. As mentioned, the performance in the replay-run can be significantly improved through reducing the number of execution orders directly recorded.

\begin{figure}[htbp!]
\includegraphics[width=0.9\textwidth]{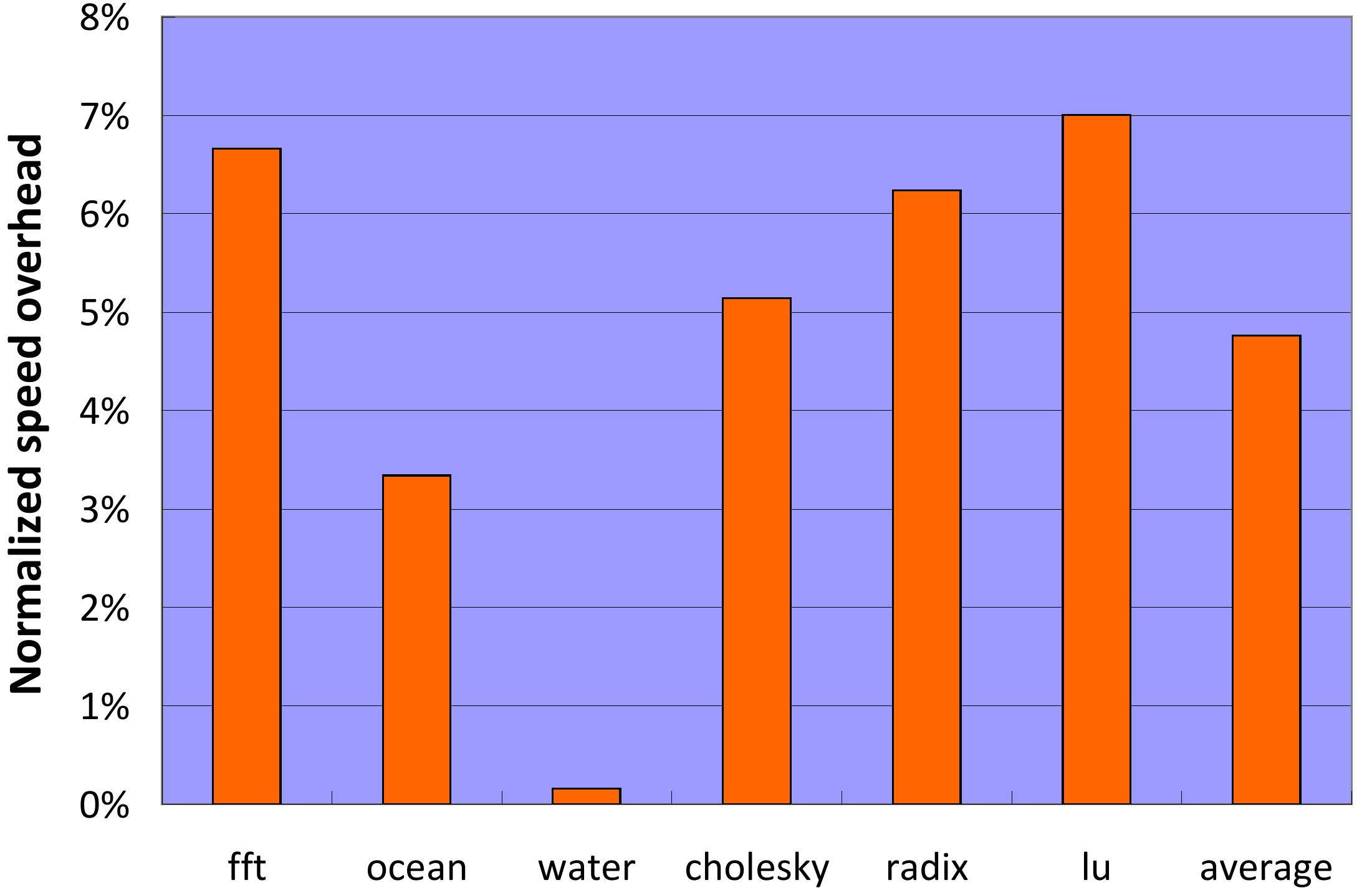}
\caption{Normalized speed overhead for 8 threads in the replay-run.} \label{fig:14}
\end{figure}

We implement RepTFD on the RTL design of the 8-core Godson-3, and carry out experiments over SPLASH2 benchmarks to validate the efficiency and effectiveness of RepTFD. Table \ref{tab:Godson-3} lists the detailed features of Godson-3 as \cite{Chen2010}. In the experiments, when we mention ``the performance of RepTFD'', we refer to the performance of redundantly executing an 8-thread application on 16-core CMP (the first-run and the replay-run of the application are simultaneously executed on 8 checked cores and 8 redundant cores respectively). When we mention ``the baseline performance'', we refer to the performance of executing an 8-thread application on 16-core CMP solely. Figure \ref{fig:14} presents the performance of RepTFD normalized to the baseline performance with 512-cycle sampling span. The average performance overhead of is 4.76\% over benchmarks\footnote{As references, two state-of-the-art transient fault detection schemes Reunion \cite{Smolens2006} and DCC \cite{LaFrieda2007} have $5\%-250\%$ and $19.2\%$ performance overheads for 16-core systems (8 checked cores + 8redundant cores) respectively.}. For benchmark water, the performance overhead is even as low as $0.16\%$.

%As a result, RepTFD only pays 3.25\% performance overhead to enforce the 1\% non-inferable execution orders and the performance loss when another 8-thread application are executed simultaneously on the 16-core processor.

%In the first-run, RepTFD incurs no performance overhead since the normal functionality of Godson-3 are not affected. As a result, the overhead of RepTFD can be evaluated by the speed overhead in the replay-run (two 8-thread applications executed simultaneously on a 16-core Godson-3).

\begin{figure}[htbp!]
\includegraphics[width=0.9\textwidth]{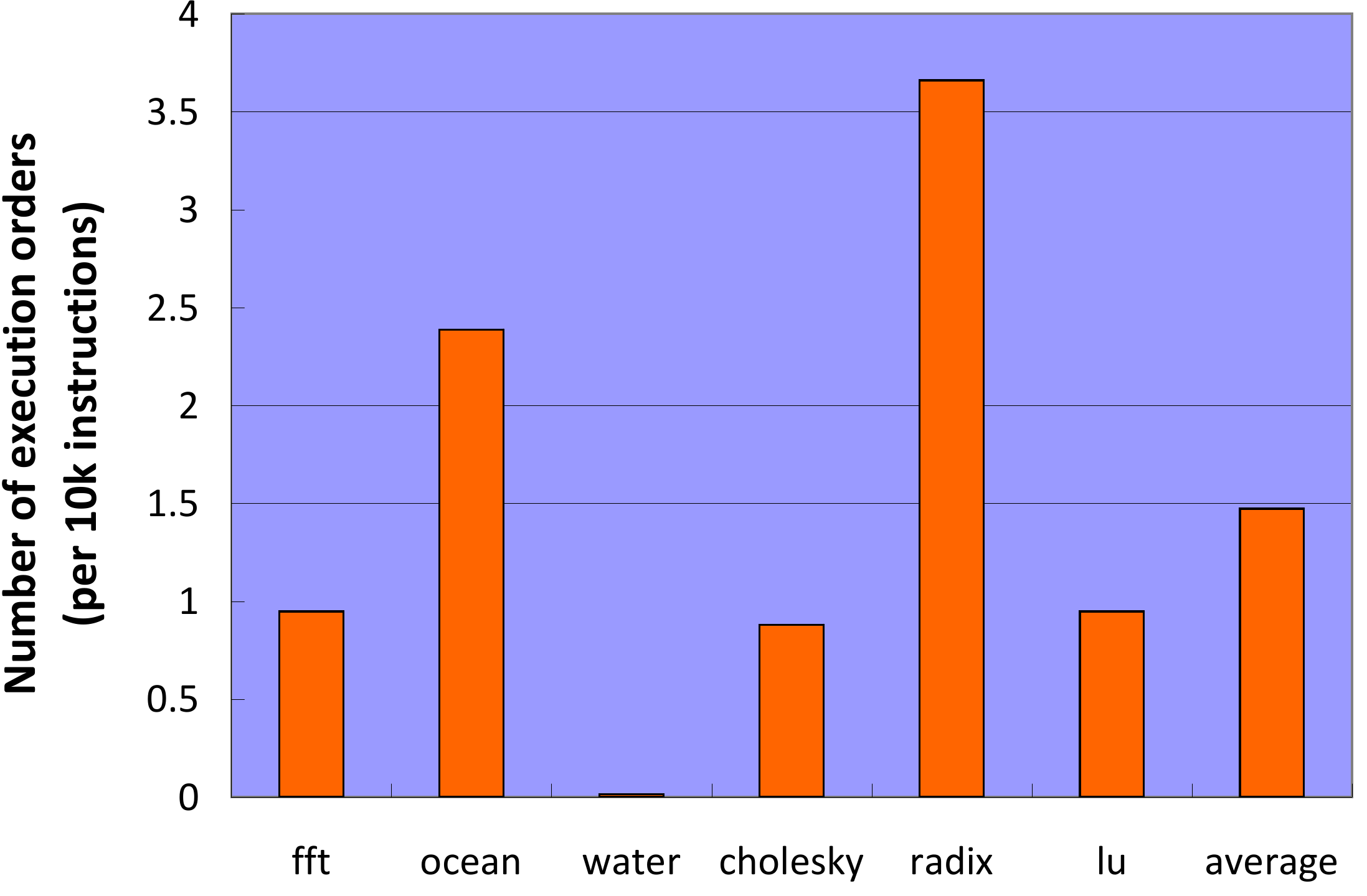}
\caption{Number of non-inferrable execution orders (per 10k instructions in one thread).} \label{fig:111}
\end{figure}

In general, the low performance overhead of RepTFD comes from two factors: the low cost to enforce non-inferrable execution orders, and the low cost to enforce physical time orders. RepTFD pays negligible cost to enforce non-inferrable execution order, since there is quite few non-inferrable execution orders that need to be enforced in the replay-run. Figures \ref{fig:111} shows the number of non-inferrable execution orders per 10000 instructions (per thread). Averagely, only 1.47 execution orders should be enforced for every 10000 instructions. For benchmark \emph{water}, only 0.014 non-inferrable execution orders per 10000 instructions should be enforced in the replay-run. Thus, negligible cost is spent on enforcing the recorded execution orders.

\begin{figure}[htbp!]
\includegraphics[width=0.9\textwidth]{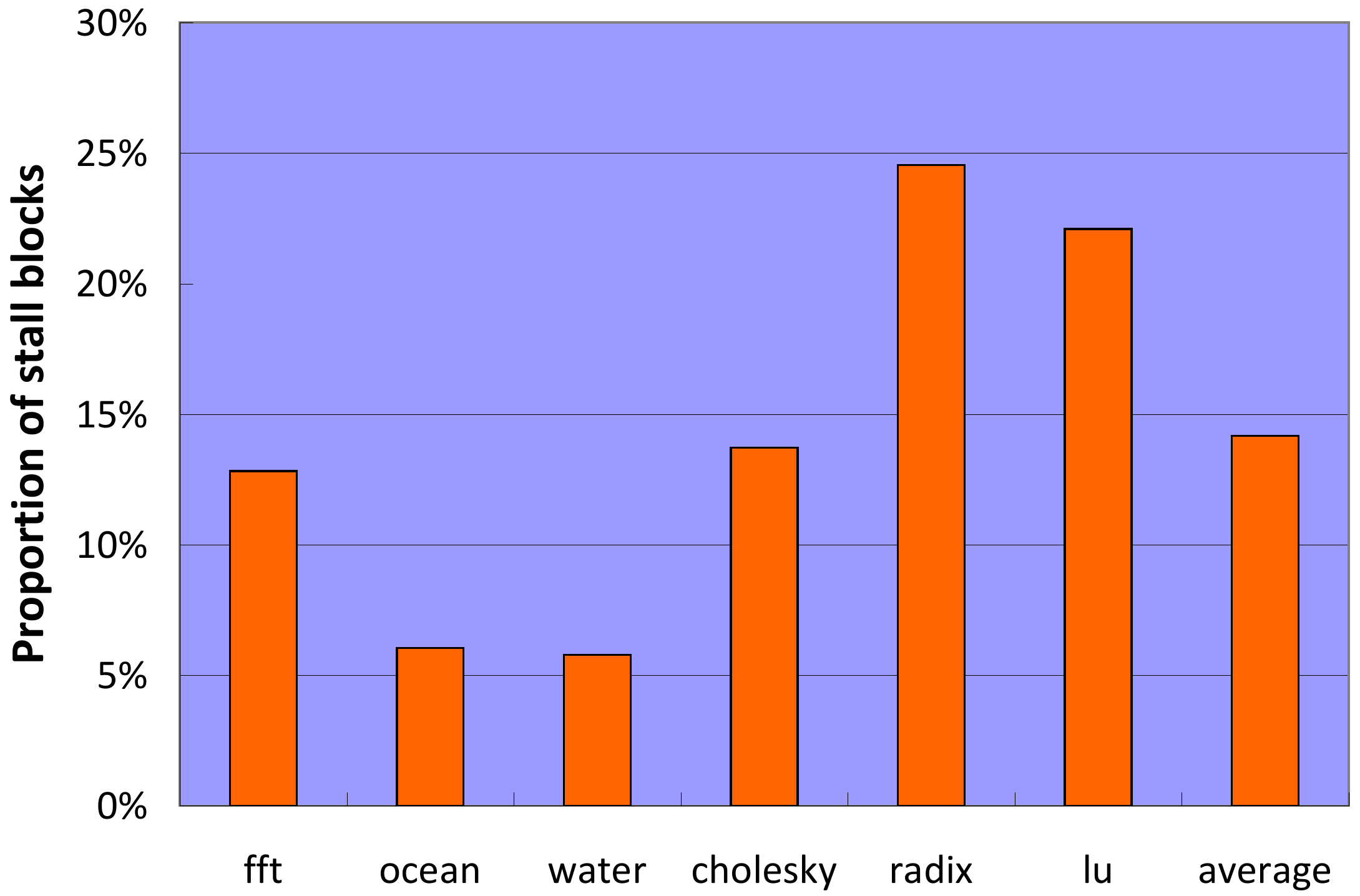}
\caption{Proportion of stalled instruction blocks.} \label{fig:15}
\end{figure}

\begin{figure}[htbp!]
\includegraphics[width=0.9\textwidth]{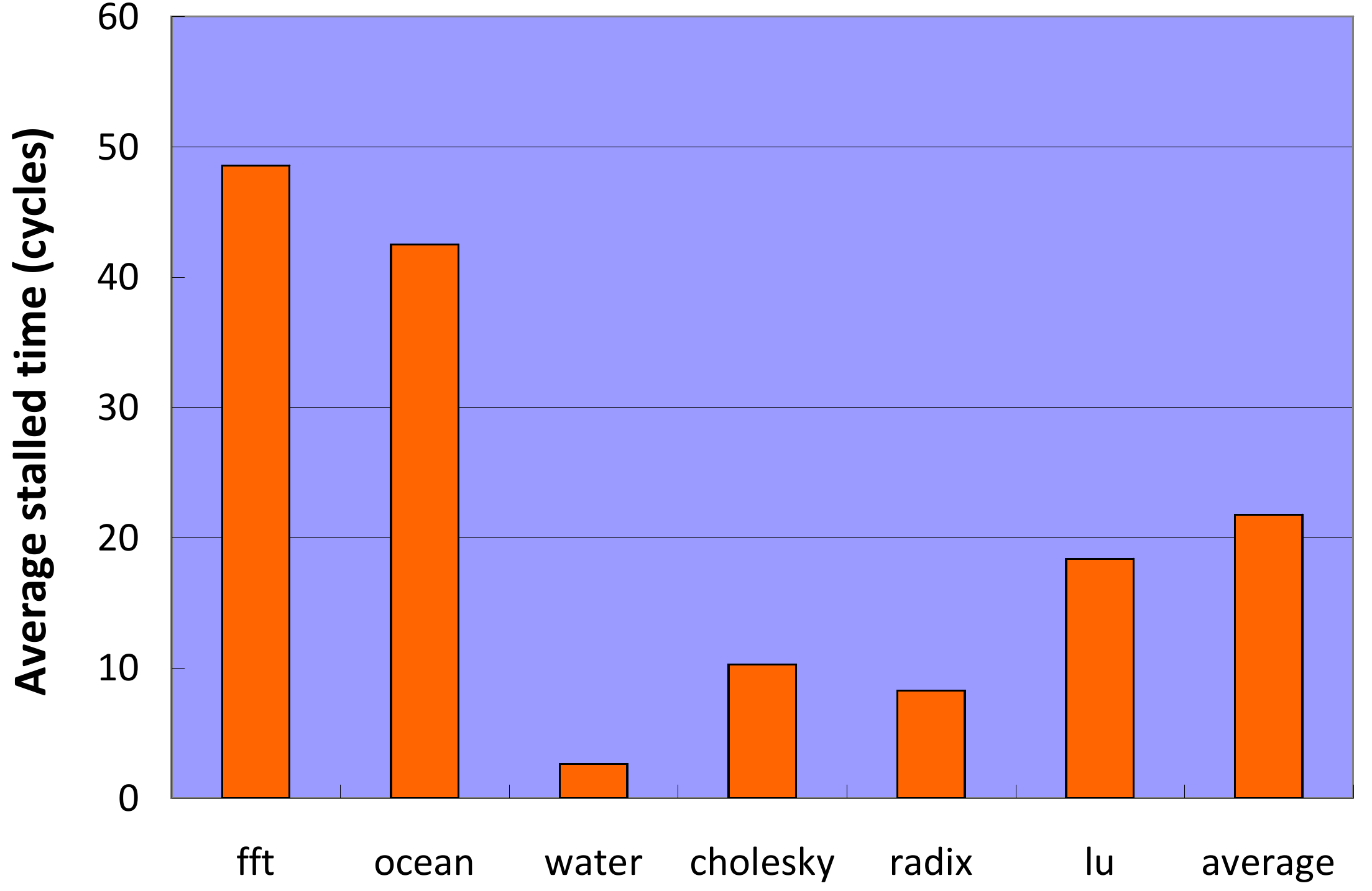}
\caption{Average stalled time for each stalled block.} \label{fig:16}
\end{figure}

On the other hand, RepTFD can cost-effectively enforce physical time orders, since the recorded pending period for each instruction block is quite relaxed (twice as the actual execution time of the block). As illustrated in Figure \ref{fig:15} and Figure \ref{fig:16}, in the replay-run, only $14.17\%$ instruction blocks are stalled to enforce the physical time orders, while for each stalled block, the average stalled time is only $21.75$ clock cycles. In other words, for all 512-cycle instruction block, we only need to stall 3.08 cycles ($14.17\% \times 21.75$) per block averagely, which brings about 0.6\% ($3.08/512$) performance penalty in the replay-run. For benchmark \emph{water}, the performance overhead is even lower: Only $5.79\%$ instruction blocks are stalled and the average stalled time for each stalled block is only $2.61$ clock cycles.


\begin{figure}[htbp!]
\includegraphics[width=0.9\textwidth]{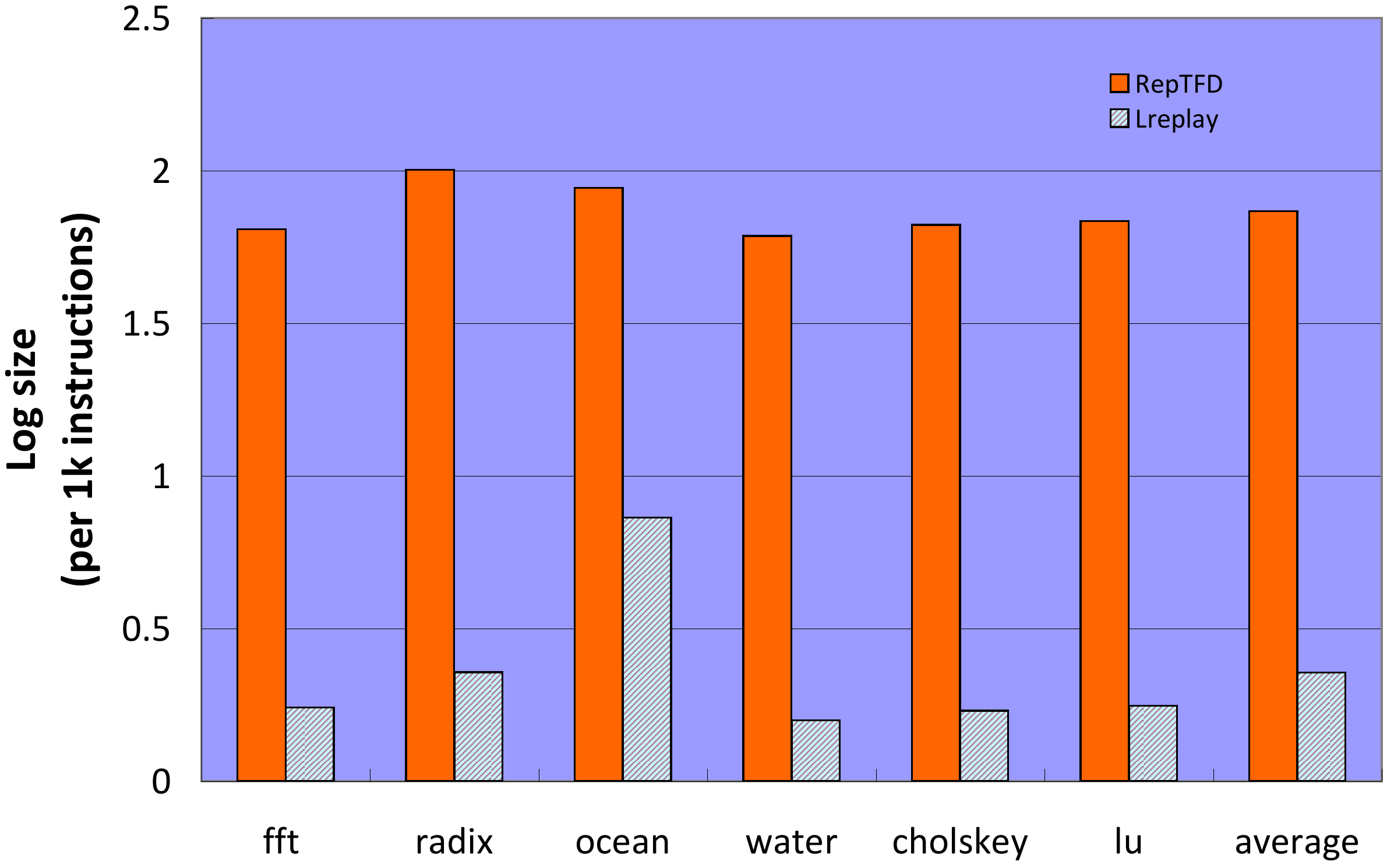}
\caption{Log size in comparison to LReplay \cite{Chen2010}.} \label{fig:17}
\end{figure}

In addition, we also evaluate the size of the determinism-log. RepTFD should record the detailed number of instructions in each block, thus has larger log size than our previous conference paper LReplay \cite{Chen2010}, which has fixed number of instructions in each block. As shown in Figure \ref{fig:17}, the average log size of this journal paper's determinism-log is 1.87 byte per kilo instructions, while LReplay's log size is only 0.36 byte per kilo instructions (with 512-cycle sampling span). However, RepTFD's log size is still acceptable for an 8-thread fault tolerant CMP: Only 14.96 MB is required to record/replay a 1-second segment of an 8-thread program (when the IPC is 1).

%Notably, since we do not replicate the uncore parts, there will be some additional performance overhead when the first-run and the replay-run are executed simultaneously as shown in Figure \ref{fig:6}. We carry out extra experiments to evaluate the overhead when two 8-thread applications are executed simultaneously on a 16-core CMP. As shown in Figure , executing two 8-thread applications simultaneously on a 16-core CMP is 5\% slower on average than one sole execution of a 8-thread application on a 16-core CMP. Thus, the performance overhead of RepTFD is \% () in total
%While for TDB, which proposes a fault-tolerant cache coherence protocol, RepTFD has a comparable performance.

Moreover, we evaluate the area consumption of RepTFD. The main area overhead of RepTFD is a $1024\times 27$ CAM per checked core, which consumes about 5.0 square millimeter as a whole, according to the report of Synopsys's Design Compiler with STMicro 65nm GP/LP mixed process. Note that the overall area of our 16-core CMP (under the same process) is about 600 square millimeter, the area consumption of RepTFD is only about 0.83\% of the whole chip.
%get the percentage of stall when enforce the orders in determinism-log. As shown in Figure \, the percentage of instruction blocks that are stalled to enforce physical time orders is . Among , which wells confirms our analysis. And the percentage of a memory instruction to wait for the execution orders is . Although most execution orders should be , there are only execution orders executed.

\section{Conclusion}\label{sec:conclusion}%(0)

In this paper, we propose RepTFD, a core-level transient fault detection scheme which employs deterministic replay to achieve 100\% detection coverage. Different from existing core-level schemes which can only protect each core separately, RepTFD protects both the cores and the uncore parts without modifications to the uncore parts of the chip. To avoid remarkable performance overhead in the replay-run, RepTFD records the relaxed pending period for each instruction block (whose length is twice of the actual execution time of the block) in the first-run. Through cost-effectively enforcing the resultant physical time orders between blocks in the replay-run, $>99\%$ execution orders, which can be inferred from physical time orders, do not need to be enforced anymore. As a result, RepTFD incurs only 4.76\% slowdown (in comparison to the normal execution without fault-tolerance).

RepTFD is the first practice to tackle transient fault detection by cutting-edge deterministic replay techniques. We uncover that through there are many previous investigations on detecting transient faults on CMPs, few of them can do this job as efficient, effective, and elegant as deterministic replay. On the other hand, the requirements aroused in transient fault detection call for a hardware-assisted deterministic replay approach different with those existing log-size-oriented replay approach. It can motivate further improvements on deterministic replay.
%lilei: 这个first attempt不合理，很多文章都有提到用重放来做故障检测的可能性。

%cy: 扯鸡巴蛋的话，主要是怕别人说我们就是用了一下确定性重放。能不能写出类似于“我们没有发明线性代数，
%但是提出用线性代数解决了某个计算机问题依然牛比”这样的感觉来？

%core-level transient fault detection scheme with 100\% coverage. In RepTFD, a parallel workload is first executed on the check group of cores, and then replayed deterministically on the redundant group of cores. By providing redundancy for a group of cores, RepTFD can catch all malignant faults in the CMP through comparing the instruction results between the two groups of cores. Besides 100\% fault coverage, the performance overhead of RepTFD is also low since only a small fraction of execution orders are directly recorded. In addition, RepTFD just adds serval registers and a CAM for each core module, while keeping most existing parts of the CMP unmodified, which indicates that RepTFD has very low area, design and verification costs.
%
%Our evaluation over SPLASH2 workloads shows that the performance overhead of RepTFD is only 3.85\% for 8 threads (16 processors). Together with 100\% transient fault coverage and low costs in performance, area and design, RepTFD shows its superiority to meet the requirements of future fault-tolerant CMPs.

\end{document}